\documentclass[10pt,a4paper]{article}

\usepackage{ndes99}
\usepackage{epsfig}
\usepackage{subfigure}

\newcommand{\scap}	{}

\newcommand{\lrb}       {\left(}        \newcommand{\rrb}       {\right)}
        
\newcommand{\beann}     {\begin{eqnarray*}}
\newcommand{\eeann}     {\end{eqnarray*}}
\newcommand{\bea}       {\begin{eqnarray}}
\newcommand{\eea}       {\end{eqnarray}}
\newcommand{\fv}        {v_{\rm F}}
\newcommand{\fk}        {k_{\rm F}}
\newcommand{\vt}        {\vartheta}
\newcommand{\vr}        {\varrho}

\newcommand{\ac}        {\mbox{\scap ac}}
\newcommand{\acm}       {\mbox{{\scriptsize {\scap ac}}}}
\newcommand{\dc}        {\mbox{\scap dc}}
\newcommand{\dcm}       {\mbox{{\scriptsize {\scap dc}}}}

\newcommand{\ii}        {{\rm i}}
\newcommand{\ee}        {{\rm e}}

    \title{Anomalous Tien--Gordon scaling in a {\scap 1d} tunnel junction}
\name{G. Cuniberti$^{1,2}$,
    A. Fechner$^{2,3}$,
        M. Sassetti$^2$, and
	    B.  Kramer$^3$}
\address{
$^1$    Max-Planck-Institut f{\"u}r Physik komplexer Systeme,
    N{\"o}thnitzerstra{\ss}e 38, D-01187 Dresden\\
    $^2$    Dipartimento di Fisica, INFM,
        Universit{\`a} di Genova,
	    via Dodecaneso 33, I-16146 Genova\\
	    $^3$    I. Institut f\"ur Theoretische Physik,
	        Universit{\"a}t Hamburg,
		    Jungiusstra{\ss}e 9, D-20355 Hamburg \\
{\small e-mail: {\tt cunibert@mpipks-dresden.mpg.de} url: {\tt www.mpipks-dresden.mpg.de/\symbol{126}cunibert}}
}

\begin{document}

    \maketitle
    
\begin{abstract}
\noindent We investigate the nonlinear \ac \ transport through a quantum wire
    with an impurity in the presence of finite range
    electron--electron interactions.
    We discuss the influence of the spatial shape of the \ac \ electric field
    onto transport properties of the system and find that
    the scaling behavior of the occupation probability of the 
    sidebands depends on the range of the voltage drop.
\end{abstract}

\section{Introduction}

Time dependent quantum transport has attracted a lot of
interest since the works of Tien and Gordon \cite{TG63} and Tu\-cker
\cite{Tucker79}; more recently, theoretical findings \cite{WJM93,BS94} and experiments on quantum dots \cite{KJOMcENMS94} and on superlattices \cite{WJZA97} renewed the interest in photon--assisted transport in semiconductor nanostructures. 
In particular, the possibility to investigate experimentally 
time--dependent transport th\-rou\-gh mesoscopic systems has opened the way to a
deeper understanding of new effects strongly relying on the spatiotemporal
coherence of electronic states. 
Moreover, in most time--dependent experiments like electron pum\-ps
\cite{GAHMPEUD90,KJvdVHF91}, photon--assisted--tunneling
\cite{KJOMcENMS94,OFvdWIHTK98,KAGKCGBR95},
and lasers \cite{FCSSHC94} require an analysis going beyond the linear response
theory in the external frequency.
Thus, many efforts have been devoted, in last years, to
the theoretical investigation of nonlinearities in semiconductor nanostructures
\cite{LGTP98,ZKKP96}, electronic correlations \cite{AP98,JWM94}, and
screening of \ac \ fields \cite{CSK98-prb,PB98}. 

The Tien--Gordon formula, according to which the \dc \ component of the
photo--induced 
current is given by a superposition of static currents $I_0$ (the currents
without the \ac \ field) weighted by integer order Bessel functions, is
represented by the following formula 
    \bea
    I_{\dcm} =
    \sum_{n=-\infty}^\infty J_n^2 \lrb  \frac{eV_1}{\hbar \Omega} \rrb
	       I_0 \lrb V_0 + n \hbar \Omega/ e \rrb ;
	       \label{eq:TG}
    \eea
the argument of the Bessel functions is linearly  dependent on the \ac \ voltage intensity $V_1$ and on the inverse of the driving frequency (or subharmonic) $\Omega$.  
A selfconsistent theory, based on the scattering matrix approach, has shown
that the side--band peaks depend on the screening properties of the system
\cite{PB98};
moreover theoretical investigations for superlattice microstructures showed an $\Omega^{-2}$
dependence of the transmission probability spectrum of the photonic sidebands (that is the argument of the Bessel functions), when a nonlocalized (a finite range) \ac \ driving was taken into account \cite{Wagner96,WZ97,Wagner98}.

In this paper, we investigate how {\scap 1d} electron--electron interaction, in the framework of the Luttinger model \cite{Haldane81a,Haldane81b,Voit95}, nonlinearities, due to the presence of an impurity,
and a finite range  \ac \ electric field affect the photo--induced current. 
We will show that the TG formula is still valid,
but the argument of the Bessel functions is not anymore linearly dependent on $1 / \Omega$.


In the time dependent regime the nonlinearity of the system gives rise to
frequency mixing  and harmonic generation.
Earlier treatments of the \ac \ transport considered
voltages, dropping only at the position of the barrier \cite{SWK96,FSK99},
and zero range interactions between the electrons.
Here, both of these are generalized to the more realistic situation of finite range of both, the electron--electron interaction and the electric field.
As a matter of fact previous calculations \cite{SK96}
showed clearly that the spatial
shape of the electric field does influence \ac \ transport.


\section{Model}

    The Hamiltonian for a Luttinger liquid of length $L$ ($\to \infty$)
    with an impurity and subject to a time--dependent electric 
    field is 
    $H = H_0 + H_{\rm imp} + H_{\acm}$, where
    \bea
    H_0 = \sum_{k \neq 0} \hbar \omega(k)^{\phantom{\dagger}} 
    b_k^\dagger b_k^{\phantom{\dagger}}.
    \label{H0} 
    \eea
    The dispersion relation of the collective excitations, 
\beann
    \omega_k =\fv |k|  \sqrt{1 + \hat{V}_{\rm ee} (k) / \hbar \pi \fv},
\eeann
    depends 
    on the Fourier transform of the finite range interaction potential
    \cite{CSK98-prb}.
    We assume a {\scap 3d} screened Coulomb potential of range $\alpha^{-1}$ 
    projected 
    onto a quantum wire of diameter $d \approx \alpha^{-1}$. 
The interaction
    decays exponentially and one gets
    $V_{\rm ee}(x)=(V_{\rm L} \alpha /2) \ee^{- \alpha |x|}$,
    with interaction strength $V_{\rm L}$
    \cite{CSK96-jpcm}. 
    For $\alpha \rightarrow \infty$,
    one obtains a zero--range interaction.

    The tunneling barrier of height $U_{\rm imp}$ is localized at $x=0$
    \cite{KF92a,KF92b},
    \bea
    H_{\rm imp} = U_{\rm imp} \cos \lrb 2 \sqrt{\pi} \vt (x=0 ) \rrb
    \label{Himp},
    \eea
    with the phase variable of the Luttinger model
    \beann
    \vt (x) = \ii \sum_{k \neq 0}  {\rm sgn} (k) 
	      \sqrt{ \frac \fv {2 L \omega(k)} } \ee^{- \ii k x}
	      \lrb b_k^\dagger + b_{-k}^{\phantom{\dagger}} \rrb .
    \eeann
    The coupling to the external driving voltage
    yields 
    \beann
    H_{\acm} = e \int_{-\infty}^\infty dx \vr(x) V (x,t) .
    \eeann
    The electric field is related to the voltage drop by differentiation,
    $E(x,t) = - \partial_x V(x,t)$, and the charge
    density is $\vr (x) = \fk / \pi + \partial_x \vt (x) /\sqrt{\pi}$.
    The space--time dependent electric field,
    $E(x,t)  =  E_{\dcm} (x) + E_a (x) \cos \lrb \Omega t \rrb$, such that
    $E_{a}(x)=E_1 \ee^{-|x|/a}$,
    gives a voltage drop 
    $V_1 \equiv \int_{-\infty}^{\infty} dx E_a(x)=2 E_1 a$. 
    The spatial dependence of the \dc \ part of the electric field does not
    need to be specified, as only the overall voltage drop,
    $V_0 \equiv \int_{-\infty}^\infty dx E_{\dcm} (x)$,
    is of importance in \dc \ transport \cite{SK96}.

\section{Methods and Results}

The current at the barrier is given by the expectation value $I(x=0,t)
= \left\langle j(x=0,t)\right\rangle$, where the current operator is
defined via the continuity equation, $\partial_{x} j(x,t) = - e
\partial_t \rho(x,t)$. For a high barrier, the tunneling contribution
to the current can be expressed in terms of forward and backward
scattering rates which are proportional to the tunneling probability
$\Delta^2$.  The latter may be obtained in terms of the barrier height
$U_{\rm t}$ by using the instanton approximation \cite{Weiss93}. The
result can be written in terms of the one-electron propagator $S+{\rm
i}R$ \cite{FSK99},
\bea
\label{eq:7}
I(x=0,t)  =  e \Delta^2 \int_0^\infty {\rm d}\tau \, {\rm e}^{-S(\tau)}
    \sin R(\tau) \nonumber
    \\  \times 
    \sin\left[ 
    \frac{e}{\hbar}\int_{t-\tau}^t{\rm d}t' V_{\rm eff}(t')
    \right],  
\eea
with
\beann
\label{eq:8}
S(\tau) + {\rm i} R(\tau) = \frac {e^2}{\pi \hbar} 
    \int_0^{\omega_{\rm max}} \frac {{\rm d} \omega}{\omega} 
    {\cal R}e \left\{ \sigma^{-1} (x=0,\omega)\right\} \nonumber
    \\   \times  \left[
    (1-\cos{\omega \tau})\coth
    \frac{\beta \omega}{2} 
    + {\rm i} \sin \omega\tau
    \right],
\eeann
where $\beta = 1/k_{\rm B}T$, $\omega_{\rm max}$ the usual frequency
cutoff that corresponds roughly to the Fermi energy \cite{Solyom79}, and
the ac conductivity of the system without impurity is \cite{SK96}
\begin{eqnarray}
\label{eq:9}
\sigma (x,\omega) = \frac {-{\rm i}v_{\rm F}e^2 \omega}
{\hbar \pi^2}\int_0^\infty \!\!\!\!\!\!\!\!\!\!\!- {\rm d}k 
\frac {\cos kx}{\omega^2(k)-(\omega+{\rm i}0^+)^2}\,.
\end{eqnarray}
Furthermore, the effective driving voltage is related to the electric
field by \cite{SK96}
\bea
\label{eq:10}
V_{\rm eff}(t)&=&\int_{-\infty}^\infty{\rm d}x
\int_{-\infty}^t{\rm d}t' E(x,t') r(x,t-t') \nonumber
\\ & =& V_0 + \frac{\hbar \Omega}{e}|z|
\cos \left(\Omega t-\varphi_z\right),
\eea
where $r(x, \omega) = \sigma (x, \omega) / \sigma(x, \omega)$, $|z|$ and $\varphi_z$ are, respectively, modulus and argument of
\begin{eqnarray}
\label{eq:2}
z=\frac{e}{\hbar\Omega}\int_{-\infty}^{\infty}{\rm d}x
E_{a}(x)r(x,\Omega).
\end{eqnarray}
With the above assumptions about the
shapes of the driving field and the interaction potential one obtains
\begin{equation}
\label{eq:12}
|z|=\frac{eV_1}{\hbar
\Omega}\frac{1}{\sqrt{1+a^2 k^2(\Omega)}} A \left(\frac{\Omega}{v_{\rm F}
\alpha}, \frac{k(\Omega)}{\alpha}, \alpha a\right),
\end{equation}
where $k(\Omega)$ is the inverse of the dispersion relation and
\begin{equation}
\label{eq:13}
A^2\left(u,v,w \right) 
= \frac{1}{1+u^2}\left[1+v^2\frac{(u+wv)^2}{(uw+v)^2}\right].
\end{equation}

In the following, we concentrate on the results for the dc component
of the current which does not depend on $x$ and is directly given by
the current at the barrier, for which we only need to know only $|z|$,
\begin{eqnarray}
\label{eq:14}
I_{\rm dc} 
= \sum_{n=-\infty}^\infty J_n^2 \left(|z|\right)I_0\left(V_0+n
\frac{\hbar\Omega}{e}\right).         
\end{eqnarray}

The important point here is that the driven dc current is completely
given in terms of $I_0(V_0)$, the nonlinear dc current-voltage
characteristic of the tunnel barrier,
\begin{eqnarray}
\label{eq:15}
I_0 \left(V_0\right)=e\Delta^2\int_0^\infty \!\!\!\! {\rm d}\tau
{\rm e}^{-S(\tau)} \sin R(\tau) 
\sin \left(\frac{eV_0 \tau}{\hbar}\right).
\end{eqnarray}
Eqs. (\ref{eq:14}), (\ref{eq:15}) generalize results which have been
obtained earlier \cite{TG63} but {\em without} interaction between
the tunneling objects, and also for the Luttinger model with a
zero-range interaction, together with a $\delta$-function like
dri\-ving electric field \cite{SWK96}.

For $V_0$ much smaller than some cutoff-voltage $V_{\rm c}$ whi\-ch is
related to the inverse of the interaction range, $I_0 \propto
V_0^{2/g-1}$. This recovers the result obtained earlier for
$\delta$-function interaction and zero-range bias electric field
\cite{KF92a}. When $V_0\gg V_{\rm c}$, the current becomes linear
\cite{SK99}. For intermediate values of $V_0$, $I_0$ exhibits a
cross-over between the asymptotic regimes with a point of inflection
near $V_{\rm c}$.  For zero-range interaction, $I_0 \propto
V_0^{2/g-1}$ for any $V_0$. 
    
\begin{figure}[h]
\begin{center}
\subfigure{\epsfig{file=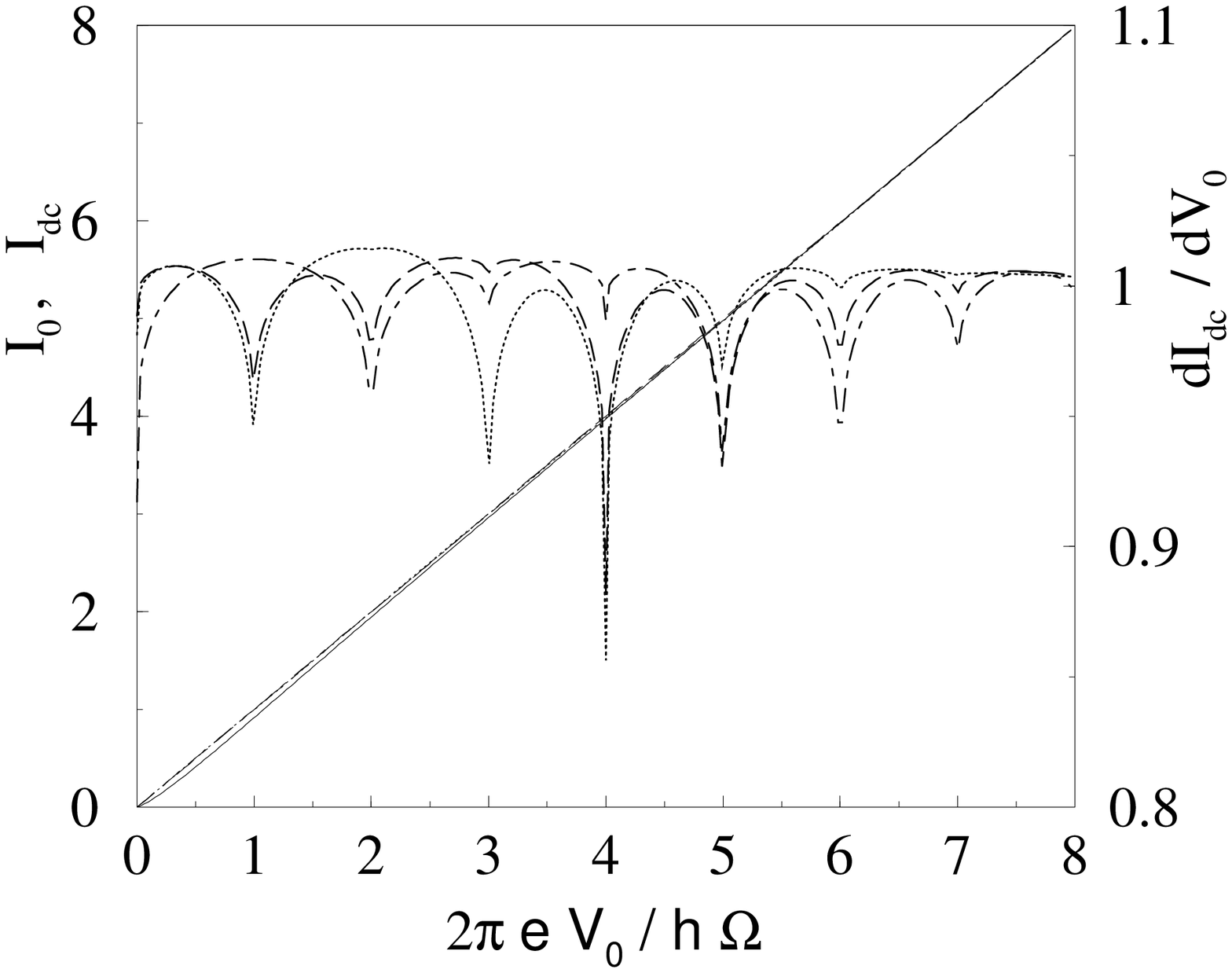, width=16pc} }
\subfigure{\epsfig{file=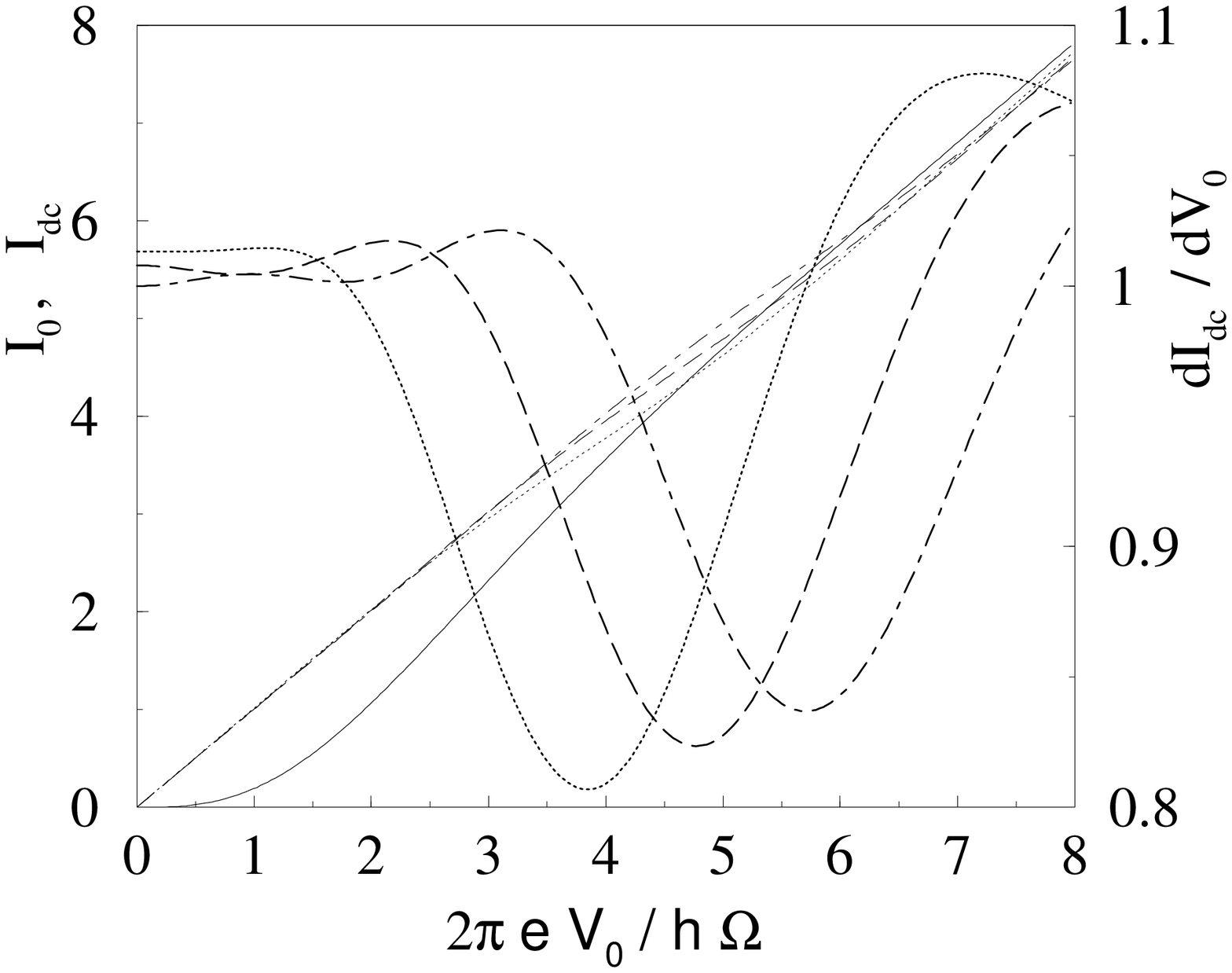, width=16pc} }
\end{center}
 \caption[1]{Currents $I_0$, $I_{\rm dc}$ and differential 
   conductance ${\rm d}I_{\rm dc}/{\rm d}V_0$ at zero temperature
   as a function of the ratio $e V_0/\hbar \Omega$ for $g=0.9$
   (top), $g=0.5$ (bottom) for values $\Omega= v_{\rm F} \alpha$,
   $a=0$, and $e V_1/ \hbar v_{\rm F} \alpha =\ell$ ($\ell=5$
   dotted, $\ell=6$ dashed, $\ell=7$ dash-dotted lines). Currents
   in units of $\hbar v_{\rm F} \alpha / e R_{\rm t}$;
   differential conductance in units of $R^{-1}_{\rm t}$;
   tunneling resistance $R_{\rm t}=2 \hbar\omega^2_{\rm max}/ \pi
   e^2 \Delta^2$.}
 \label{fig:1}
\end{figure}

Figure~\ref{fig:1} shows the currents $I_0$, $I_{\rm dc}$ and the
differential conductance ${\rm d}I_{\rm dc}/{\rm d}V_0$ as functions
of $eV_0/\hbar\Omega$ for $g=0.9$ and $g=0.5$ for zero-range of the
driving electric field.  For $g=0.9$ one observes sharp minima in the
differential conductance at integer multiples of the driving
frequency in certain regions of the driving voltage $V_1$. These can
be understood as follows. When the strength of the
interaction is not too large, the region where ${\rm d}I_{\rm
dc}/{\rm d}V_0$ is much smaller than 1 is small compared with
$\hbar\Omega$ thus for $eV_0\approx \hbar\Omega$, ${\rm d}I_{\rm
dc}/{\rm d}V_0\propto
\left(2/g-1\right)|eV_0-\hbar\Omega|^{2/g-2}$. Then, Eq.
(\ref{eq:14}) yields near $eV_0=m\hbar\Omega$
\begin{eqnarray}
\label{eq:16}
\frac{{\rm d}I_{\rm dc}}{{\rm d}V} &\approx& 1-J_m^2(|z|)+
{\rm const}\cdot
J_m^2(|z|) 
\nonumber
\\ & & \times \left|eV_0-m\hbar\Omega\right|^{2/g-2}.
\end{eqnarray}
For $g>2/3$, this yields for integer $m$ the cusp-like structures
observed in Fig.~\ref{fig:1}. For $g<2/3$, no cusps occur anymore. In
addition, the current $I_{\rm dc}$ is depleted so strongly and over
such a large region of the bias voltages that the regime of almost
vanishing ${\rm d}I_{\rm dc}/{\rm d}V_0$ becomes larger than
$\hbar\Omega$ and in general no minima near integer multiples of the
frequency exist. As can be seen in the figure, the depths of the
cusps depend on the driving voltage $V_1$ ($\propto |z|$) which can
also be understood from of Eq.~(\ref{eq:16}) which shows that the
values of the differential conductances at the voltages
$eV_0=m\hbar\Omega$ are approximately $1-J_m^2(|z|)$.

\begin{figure}[h]
\begin{center}
\epsfig{file=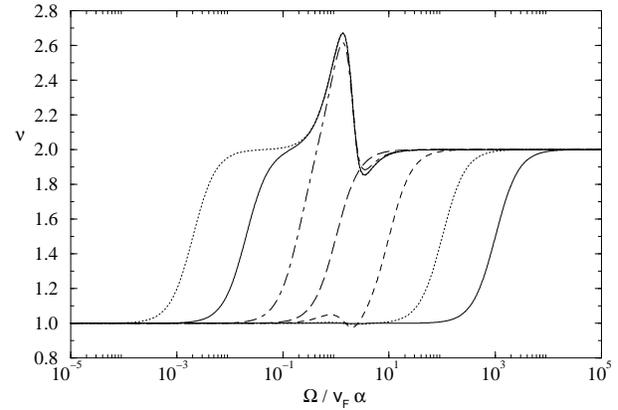, width=\linewidth}
\end{center}
 \caption[2]{Scaling exponent $\nu$ of the argument $|z|$ of the Bessel
   functions as a function of $\Omega$, for ranges of the driving
   field (curves from right to left) $\alpha a = 10^{-3}$,
   $10^{-2}$, $10^{-1}$, $1$, $10$, $10^{2}$, $10^{3}$, and
   $g=0.5$.}
 \label{fig:2}
\end{figure}

It is therefore instructive to look into the behavior of $|z|$ as a
function of the frequency. Figure \ref{fig:2} shows the scaling
exponent $\nu$ determined from
\begin{eqnarray}
\label{eq:17}
\nu = -v_{\rm F} \alpha \frac{{\rm d} \log |z|}
{{\rm d} \log \Omega}.
\end{eqnarray}

We observe a non-universal cross-over between $|z|\propto
\Omega^{-1}$, the case discussed by Tien and Gordon \cite{TG63} which
corresponds to a driving field of zero-range ($a\to 0$), and
$|z|\propto \Omega^{-2}$ which is obtained for a homogeneous external
field ($a\to\infty$) \cite{Wagner96}. Although the behavior of $z$ depends
strongly on the parameters of the model in the cross--over regime,
this does not influence qualitatively the occurrence of the cusps.
Their existence depends crucially on the finite range of the
interaction, and the condition $g>2/3$. However, by varying $|z|$,
the depths of the minima are changed due to the variation of
$J_m^2(|z|)$.

Finally, we have demonstrated that the result which has been obtained by Tien
and Gordon for tunneling of non-interacting quantum objects in 1D
driven by a mono-chro\-ma\-tic field localized at the tunnel barrier
remains valid even in the presence of interactions of arbitrary range
and shape, and for an arbitrary shape of the mono-chromatic driving
field. The central point is that the frequency driven current is
completely given by a linear superposition of the current-voltage
characteristics at integer multiples of the driving frequency,
weighted by Bessel functions.

The argument of the latter contains the amplitude of the driving
voltage only linearly but the dependence of the argument on the
frequency and the range of the driving field is determined by its
spatial shape. However, one can easily identify regions where the
dependence on the frequency becomes very simple. For a driving field
which is localized near the tunnel barrier, the integral in
Eq.~(\ref{eq:2}) can be evaluated approximately by noting that
$r(x,\Omega)$ varies only slowly with $x$ and can be taken out of the
integral. Then, $|z|=e V_1/ \hbar \Omega$ which corresponds to the
result of Tien and Gordon \cite{TG63}. In the other limit of an
almost homogeneous electric field, $E_1=V_1/a$, one needs to
calculate the spatial average of $r(x,\Omega)$ \cite{SK96}. This
gives $\sigma(k=0,\Omega)/\sigma(x=0,\Omega)\approx \Omega^{-1}$,
since $\sigma(x=0,\Omega)\approx {\rm const}$. This implies $|
z|\propto\Omega^{-2}$. Such a frequency dependence has been
discussed earlier for non-interacting particles \cite{Wagner96}. Here, we
see that it is valid under quite general assumptions also for
interacting particles. A possible method to detect this behavior
experimentally is to investigate the real part of the first harmonic
of the current through the tunnel contact and to determine the {\em
current responsivity} which is given  by the ratio of
the expansions of $I_{\rm dc}$ and the first harmonic to second and
first order in $|z|$, respectively \cite{Tucker79}.

Given the above result for the driven dc-current, the general
behavior of the differential conductance as a function of
$eV_0/\hbar\Omega$ can be straightforwardly obtained. Of special
interest is the occurrence of cusps at $eV_0/\hbar\Omega=m$ ($m$
integer) which appear to be quite stable against changes in the model
parameters. A similar result has been discussed earlier \cite{LF96},
but for a small potential barrier between fractional quantum Hall
edge states which implies zero-range interaction. In the case
discussed here, the finite range of the interaction is crucial for
obtaining the cusps, due to the absence of a linear contribution
towards the current for small voltage which is characteristic of
tunneling in 1D dominated by interaction. The cusps could be used to
frequency-lock the dc part of the driving voltage.

To summarize, we have shown how the electron correlation and the spatial
distribution of a driving field determine the anomalous scaling of the
photo--induced
current and the mode locking patterned structure of the nonlinear differential
conductance.

This work has been supported by EU via TMR(FMRX-CT96-0042, FMRX-CT98-0180), by INFM via \\ PRA(QTMD97), and by italian ministry of university via \\ MURST(SCQBD98).

{\small    

}
\end{document}